\documentclass[conference]{IEEEtran}
\IEEEoverridecommandlockouts

\usepackage{graphicx}
\usepackage{xcolor}

\usepackage{cite}
\usepackage{url}
\usepackage{hyperref}

\usepackage{array, amsmath, amssymb, amsfonts, bm, mleftright}
\usepackage{multirow, booktabs}
\usepackage{algorithm, algpseudocode}
\usepackage{xargs}

\def\BibTeX{{\rm B\kern-.05em{\sc i\kern-.025em b}\kern-.08em
    T\kern-.1667em\lower.7ex\hbox{E}\kern-.125emX}}
\begin{document}

\setlength\abovedisplayskip{1.5mm}
\setlength\belowdisplayskip{1.5mm}

\title{
Formula-Supervised Sound Event Detection:\\ Pre-Training Without Real Data \\
\thanks{This work is based on a project,
JPNP20006, commissioned by NEDO. The computational resource of ABCI provided by AIST was used.}
}

\author{
    \IEEEauthorblockN{
        Yuto Shibata\IEEEauthorrefmark{1}\IEEEauthorrefmark{2}, %
        Keitaro Tanaka\IEEEauthorrefmark{3}\IEEEauthorrefmark{2}, %
        Yoshiaki Bando\IEEEauthorrefmark{2}, %
        Keisuke Imoto\IEEEauthorrefmark{2}\IEEEauthorrefmark{4}, %
        Hirokatsu Kataoka\IEEEauthorrefmark{2}\IEEEauthorrefmark{5}, %
        and Yoshimitsu Aoki\IEEEauthorrefmark{1}\vspace{.0\baselineskip}}%
    \IEEEauthorblockA{
        \IEEEauthorrefmark{1}Keio University, Japan
        \IEEEauthorrefmark{2}National Institute of Advanced Industrial Science and Technology (AIST), Japan
    }
    \IEEEauthorblockA{
        \IEEEauthorrefmark{3}Waseda University, Japan
        \IEEEauthorrefmark{4}Doshisha University, Japan
        \IEEEauthorrefmark{5}University of Oxford, United Kingdom
    }
}

\maketitle

\begin{abstract}
In this paper, we propose a novel formula-driven supervised learning (FDSL) framework for pre-training an environmental sound analysis model by leveraging acoustic signals parametrically synthesized through formula-driven methods. Specifically, we outline detailed procedures and evaluate their effectiveness for sound event detection (SED).
The SED task, which involves estimating the types and timings of sound events, is particularly challenged by the difficulty of acquiring a sufficient quantity of accurately labeled training data. Moreover, it is well known that manually annotated labels often contain noises and are significantly influenced by the subjective judgment of annotators. 
To address these challenges, we propose a novel pre-training method that utilizes a synthetic dataset, Formula-SED, where acoustic data are generated solely based on mathematical formulas.
The proposed method enables large-scale pre-training by using the synthesis parameters applied at each time step as ground truth labels, thereby eliminating label noise and bias.
We demonstrate that large-scale pre-training with Formula-SED significantly enhances model accuracy and accelerates training, as evidenced by our results in the DESED dataset used for DCASE2023 Challenge Task 4. The project page is at \url{https://yutoshibata07.github.io/Formula-SED/}.
\end{abstract}

\begin{IEEEkeywords}
sound event detection, pre-training without real data, environmental sound synthesis
\end{IEEEkeywords}

\section{Introduction}
Sound event detection (SED)~\cite{mesaros2021sound, cakir2017convolutional} is a task that aims to estimate the acoustic events' types and their onset/offset timestamps. SED has diverse applications, including anomaly detection~\cite{surveillance} and smart home systems~\cite{turpault2019sound}. Model training and evaluation often use weak (clip-level) or strong (frame-level) labels. 
Numerous prior studies in acoustic scene analysis have pointed out data collection difficulties due to the high annotation cost~\cite{chen2022beats,weaklabeltransfer}. This is critical in SED, as it predicts the strong labels (the timestamps of event occurrences) created by human annotators. Such detailed annotations make data collection labor-intensive and expensive, hindering the development of high-resolution sound analysis systems.

In SED, where the collection of frame-level labels is challenging, weakly supervised learning that utilizes clip-level labels and self-supervised learning has been explored~\cite{weak_supervised, serizel2018large, weaklabeltransfer, chen2022beats}. For example, methods have been proposed that incorporate the Acoustic Spectrogram Transformer~\cite{gong2021ast}, pre-trained on an audio tagging task with weak labels~\cite{gemmeke2017audio}, into a SED system~\cite{ast-sed,fine-tuni-atst}. Additionally, BEATs~\cite{chen2022beats} employs self-supervised learning through patch masking and discrete label prediction to acquire semantic-rich representations. These methods have demonstrated high performance on the DESED dataset~\cite{turpault2019sound, kim2023semi, chen2023dcase}. However, since strong labels are not used during the pre-training, these methods may not be fully optimized for SED tasks that require high temporal resolution. 
Additionally, audio data contains extensive information about individuals' identities and their environments, raising privacy concerns~\cite{nelus2021privacy, liang2020characterizing, Shibata_2023_CVPR, hearyouraction}. Along with issues related to data ownership, the large-scale collection of real-world audio data still presents significant challenges.

\begin{figure}[t]
    \centering
    \includegraphics[width=0.9\hsize]{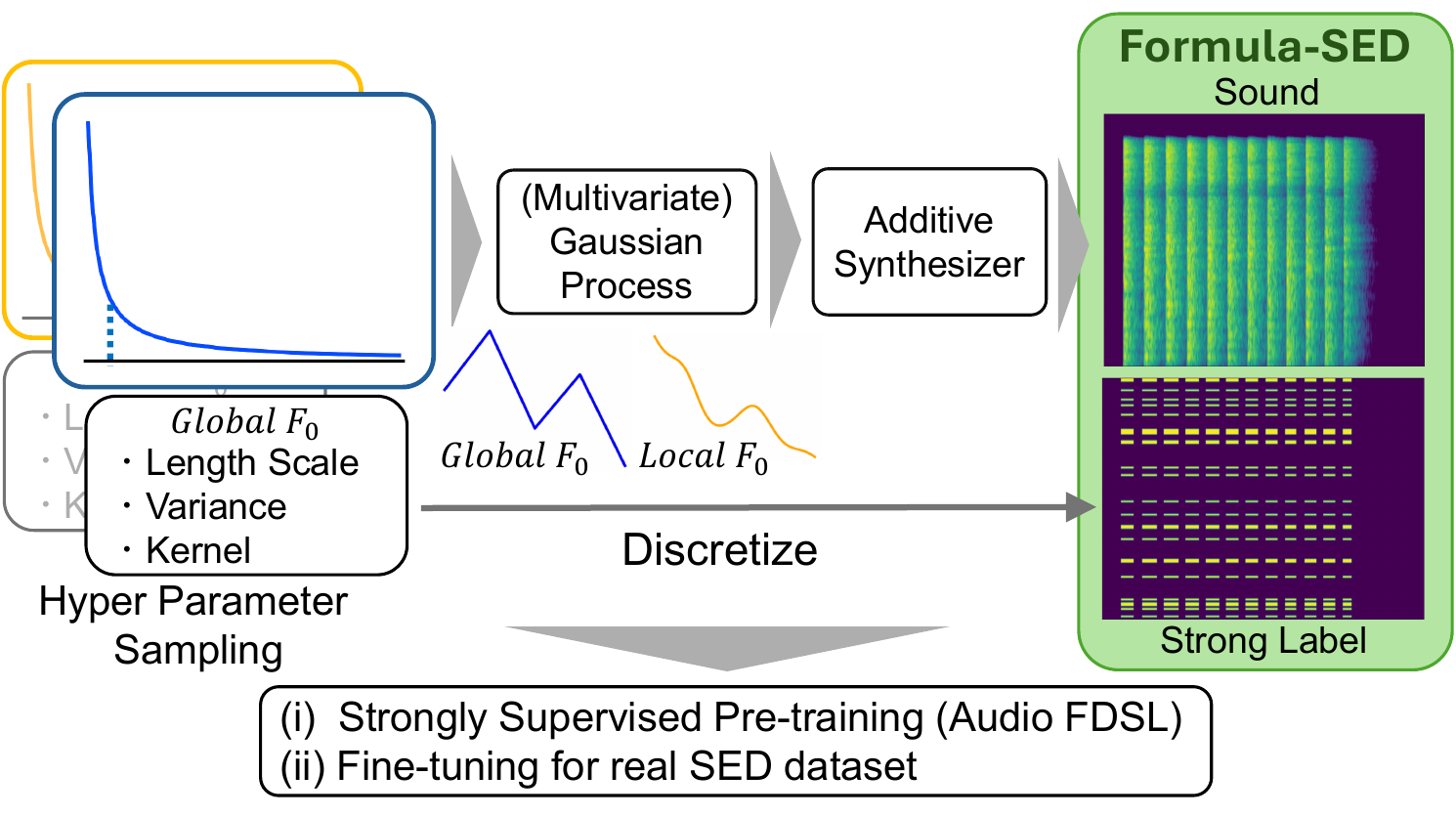}
    \vspace{-3mm}
    \caption{The overview of our proposed method. We effectively pre-train SED models using acoustic data generated solely based on mathematical formulas.}
    \label{fig:overview}
    \vspace{-5mm}
\end{figure}

In this study, we propose a method for large-scale strongly supervised pre-training of acoustic analysis models without using any real data (Fig.~\ref{fig:overview}). In the field of computer vision, it has been demonstrated that formula-driven supervised learning (FDSL), which uses fractal images and their generation parameters as labels, can achieve high performance without relying on real data during pre-training~\cite{kataoka2020pre, Nakashima_arXiv2021, kataoka2022replacing, anderson2022improving, Shinoda_2023_ICCV, nakamura2024scaling}. 
Similarly, we create a synthetic dataset for SED, named Formula-SED, and propose a novel formula-driven pre-training method that uses acoustic synthesis parameters as labels with correct timestamps.

\begin{figure}[t]
    \centering
    \includegraphics[width=1.0\hsize]{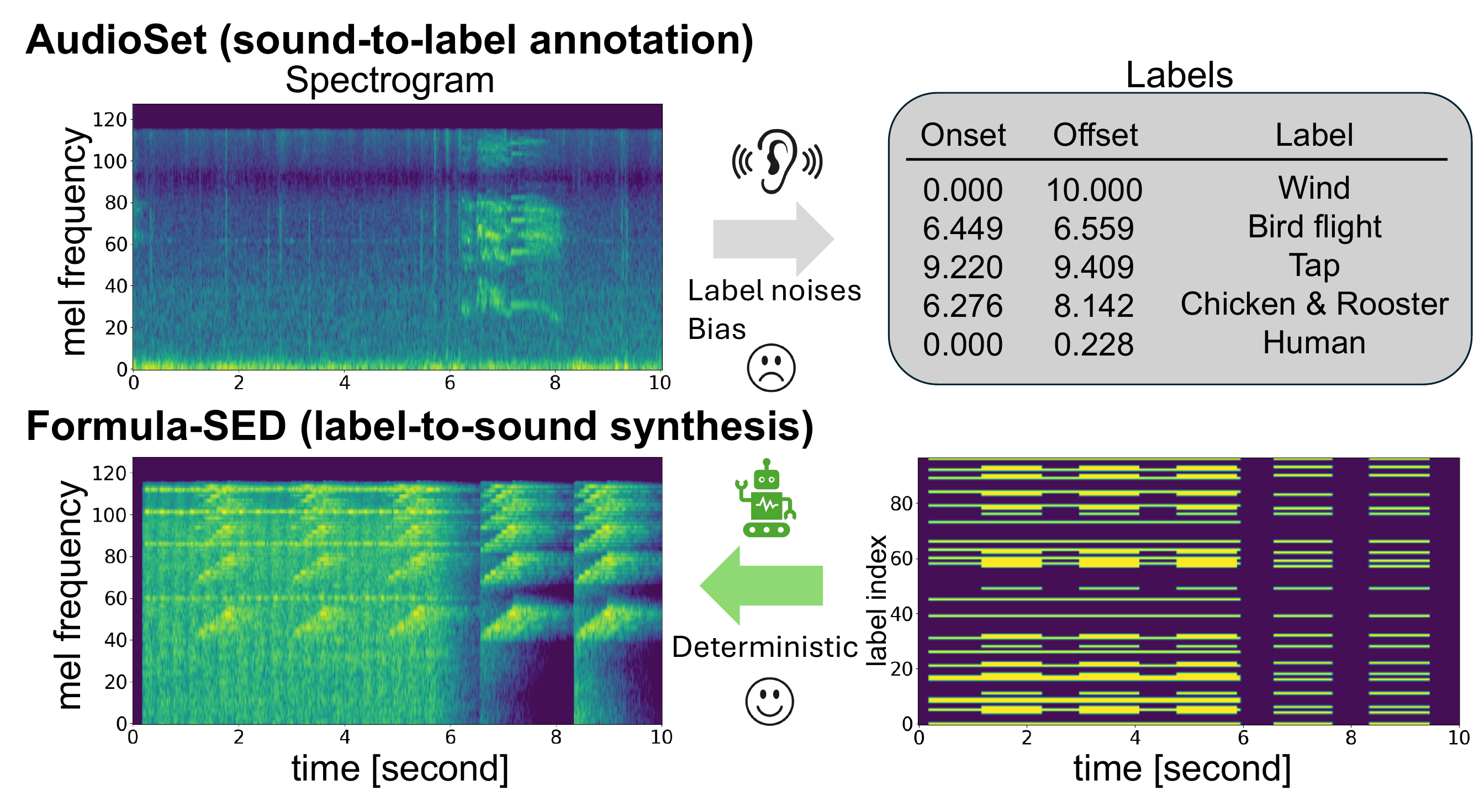}
    \vspace{-8mm}
    \caption{Comparison between real data (AudioSet) and our Formula-SED.}
    \label{fig:data_samples}
    \vspace{-6mm}
\end{figure}

The automatic generation of realistic acoustic data using only mathematical formulas is highly challenging. 
To synthesize acoustic pseudo-events, we sample spectral envelope, volume, and pitch sequences that vary locally and globally using Gaussian processes. To ensure coherence as acoustic events, we introduce correlations between harmonic and inharmonic components, as well as between inharmonic distributions at each time step. 
Since the labels in this dataset are deterministically generated at each time step, it eliminates label noise and bias mentioned earlier while avoiding privacy and data rights concerns (Fig.~\ref{fig:data_samples}). These high-fidelity sound and precise labels are used for large-scale supervised pre-training.

Experiments show that pre-training with the proposed dataset significantly improves both accuracy and convergence speed in SED training with real data, regardless of the model architecture, evaluation metrics, or pre-training data size.
Additionally, our parametric acoustic signal synthesis enables controlling over label components during pre-training, unlike popular high-level approaches such as masked audio prediction~\cite{chen2022beats}. Therefore, we investigate the specific characteristics of acoustic signals that are essential for acquiring transferable knowledge. Specifically, it was found that detecting frequency variations on both local and global scales during pre-training significantly enhances fine-tuning accuracy. To the best of our knowledge, this study is the first to demonstrate that mathematically generated acoustic signals yield transferable auditory representations for real-world data.

\section{Formula-Driven Acoustic Supervised Learning}
\label{sec:method}
As shown in Fig.~\ref{fig:overview}, our approach randomly samples parameters that define the characteristics of the acoustic signals, which are then used as input to a parametric acoustic synthesizer~\cite{engel2019ddsp}. These synthesis parameters serve as the ground truth labels, enabling large-scale strongly supervised pre-training.

\subsection{Parametric Synthesis for Sound Events}
\label{sec:parametric_synthesis}
In this paper, we follow the methodology outlined in~\cite{engel2019ddsp, serra1990spectral, beauchamp2007analysis}, synthesizing source signals by summing harmonic and inharmonic components and finally convolving reverberation.
At a given time sample $n$, let $A(n)$ denote the global amplitude and $c_k(n)$ and $\phi_k(n)$ represent the amplitude and phase of the $k$-th harmonic element, respectively.
The acoustic signal $x(n)$ is then obtained using an additive synthesizer as follows:
\begin{align}
    \hspace{-1mm}x(n)   = A(n) \sum_{k=1}^K c_k(n) \sin(\phi_k(n)) + F_t(n) * v(n),
\end{align}
where $v(n)\sim \text{Uniform}(-1, 1)$ represents stochastic noise.
Additionally, $*$ denotes convolution operation, and $F_t(n)$ represents a linear time-variant finite impulse response filter at time step $n$ with filter length $t$, which is used to model the inharmonic component by convolving it with $v(n)$. The phase $\phi_k(n)$ is determined using the instantaneous fundamental frequency $f_0(n)$ as follows:
\begin{align}
    \phi_k(n) = 2\pi \sum_{s=0}^n k f_0(s) + \phi_{0,k}.
\end{align}
Here, $\phi_{0,k} \in \mathcal{U}(0, 2\pi)$ represents random initial phase.
In the proposed method, the parameter functions $A(n)$, $c_k(n)$, $f_0(n)$, and $K$ for the harmonic components, as well as the parameter function $F_t(n)$ for the inharmonic components, are generated randomly. Then, hyperparameters of the distributions from which they are sampled are used as labels (see Sec.~\ref{sec:label_preparation}).

\begin{table}[t]
\centering
\caption{Synthesis Parameters and Their Number of Classes}
\vspace{-2mm}
\begin{tabular}{l|l}
\toprule
Scale & Name and number of classes \\
\midrule
Global &\textbullet~Voiced segment duration (3)\\
 & \textbullet~Harmonic volume (3) and inharmonic volume (3) \\
 &\textbullet~F0 variance (4) and bias (4) \\ 
 &\textbullet~Number of harmonics (3), envelope variance (4),\\ 
 & \hspace{2.8pt} length scale (4), sharpness (4), and kernel (8) \\
 &\textbullet~Inharmonic distribution sharpness (4), mode (10), \\ 
 & \hspace{2.8pt} and kernel (8) \\ 
 &\textbullet~Discrete or continuous pitch (2) \\ 
 &\textbullet~Reverb strength (-) \\
 \midrule
Local &\textbullet~Harmonic-Inharmonic volume correlation (2), \\
 & \hspace{2.8pt} volume variance (4) and kernel (8) \\ 
 & \textbullet~F0 variance (4), length scale (4), and kernel (8) \\ 
\bottomrule
\end{tabular}
\label{tab:synth_params}
\vspace{-4mm}
\end{table}

\subsection{Parameter Generation Using Gaussian Processes}
\label{sec:gaussian_prosses}
To synthesize diverse environmental sound, we design a sampling method for synthesis parameter functions by considering a question: ``What constitutes a single acoustic event?" 
A set of synthesis parameters used to create a single acoustic event must exhibit consistency or temporal continuity within the event. Additionally, for the harmonic and inharmonic components to originate from the same acoustic event, they must be temporally correlated. In this study, we use Gaussian processes to sample functions to represent diverse temporal changes and correlations between synthesis parameters.

Specifically, we randomly select kernels and hyperparameters from predetermined candidates or ranges and then sample parameter functions based on the Gaussian process. 
These hyperparameters are listed in Table~\ref{tab:synth_params}. 
We model variables such as the harmonic fundamental frequency and envelope using single-output Gaussian processes. On the other hand, to generate coherent acoustic signals that can be recognized as a single event, we model the global and local volumes of harmonic and inharmonic components using positive or negative correlations. Furthermore, the noise distribution in the frequency domain also has a temporal correlation. These correlations are expressed based on the intrinsic coregionalization model (ICM)~\cite{bonilla2007multi}. The parameter functions related to harmonic and inharmonic components are sampled as follows:
\begin{equation}
\begin{pmatrix}
v_{\mathrm{har}}(n) \\
v_{\mathrm{noise}}(n)
\end{pmatrix}
\sim \mathcal{GP}\left(\begin{pmatrix}
\Bar{v}_{\mathrm{har}} \\
\Bar{v}_{\mathrm{noise}}
\end{pmatrix}, \mathbf{B} \otimes \mathbf{K}(n, n')\right),
\end{equation}
where $v_{\mathrm{har}}(n)$ and $v_{\mathrm{noise}}(n)$ are functions representing harmonic and inharmonic volumes, respectively. 
Additionally, $\Bar{v}_{\mathrm{har}}$ and $\Bar{v}_{\mathrm{noise}}$ are the mean functions for the outputs, 
$\mathbf{B}$ is the coregionalization matrix, 
$\otimes$ is Kronecker product, 
and $\mathbf{K}(n, n')$ is the covariance function.
Note that by considering the correlation between harmonic and inharmonic components, we can handle not only the positive correlations that occur when these components arise simultaneously but also the negative correlations, such as those found in alternating events like speech. Correlated noise distribution in the frequency domain across multiple time steps is represented similarly. Sampled parameters are then fed into the synthesis model described in Sec.~\ref{sec:parametric_synthesis} to generate the source signals. These generated signals are mixed according to a randomly selected number of sources (up to four) to create the final audio input (Fig.~\ref{fig:data_samples}). The combination of synthesis parameter values across multiple acoustic events results in an enormous number of possibilities, greatly enhancing dataset variety.

\subsection{Ground-Truth Label Generation}
\label{sec:label_preparation}
To acquire effective auditory representations through pre-training, it is essential to prepare supervisory labels that are relevant to sound event detection. Synthesis parameters used for sound generation and our formula-driven supervised pre-training are summarized in Table~\ref{tab:synth_params}. They represent local and global characteristics of sound, including pitch, harmonic structure, and volume. Here, the parameter related to reverberation strength is only used for acoustic synthesis, as it did not improve accuracy in our preliminary evaluation.
The durations of generated signals are determined at random, with the associated acoustic labels being stored along with the corresponding timestamps.
Therefore, by using a parametric synthesizer, it is possible to automatically generate a high-quality SED pre-training dataset (Formula-SED) with both high-quality acoustic signals and accurate strong labels.

The parameters used for acoustic signal synthesis include both continuous values, such as length scale or variance, and discrete values, such as the number of harmonics. Through preliminary experiments, we found that label classification, where continuous values are discretized using predetermined thresholds, achieved higher accuracy in downstream SED tasks compared to the regression of continuous values. Consequently, in our experiments, pre-training is conducted as a multi-label classification task for each synthesis parameter. For the number of classes after labeling, please refer to Table~\ref{tab:synth_params}. 
We determine the threshold for label discretization based on intuition and the data distribution. 
We found that discretization with equal or overly fine intervals yielded suboptimal results.
However, a detailed analysis of threshold settings and accuracy is beyond this paper's scope.
Due to space limitations, the specific thresholds and kernel types are omitted in this paper. To perform predictions for these multiple labels, we represented the final label using a multi-hot vector. In this case, we create input acoustic data by summing multiple source signals. Therefore, if any sources having a specific label are included in the mixture, we activate the label (see Fig.~\ref{fig:data_samples}).

\section{Experimental Evaluation}
This section describes experiments conducted to evaluate the performance of the proposed pre-training method.

\subsection{Experimental Settings}
Following the method described in Sec.~\ref{sec:method}, we generated 1M sound samples. As baseline methods for the SED task, we adopted two variants of convolutional recurrent neural networks (CRNNs)~\cite{cakir2017convolutional}: (i) the lightweight DCASE2023 baseline model~\cite{DCASE2023Workshop}, which has 1.1M parameters, and (ii) Paderborn CRNN~\cite{ebbers2022pre}, which has 11M parameters and achieved the best results when trained with real-world weak label dataset. To facilitate a simple pre-training comparison, instead of using the forward-backward CRNN employed in previous studies~\cite{ebbers2021self, ebbers2021forward, ebbers2022pre}, we used the bidirectional CRNN. When pre-training with Formula-SED, we used 10k samples separate from the training files as validation data and applied early stopping. Also, we used data augmentation such as time-masking~\cite{park2019specaugment}, time-warping~\cite{ebbers2022pre}, time-shifting, and Mixup~\cite{zhang2017mixup}. 

To evaluate pre-training effectiveness, we addressed the DCASE 2023 Task 4~\cite{DCASE2023Workshop}. 
Here, SED models are trained using diverse annotations, including weak labels, strongly labeled synthetic soundscapes~\cite{salamon2017scaper}, and unlabeled data. They are designed to detect sound events in a domestic environment and consist of 10-second clips containing events such as alarms and human speech. Each model was trained for 200 epochs, with the final learning rates set to 0.001 for the baseline CRNN and 0.0001 for the Paderborn CRNN. 
For fine-tuning, we applied the aforementioned data augmentations, including frequency masking, and conducted mean-teacher training~\cite{tarvainen2017mean}.

We compared our pre-training method with random initialization and supervised pre-training using strong labels from AudioSet~\cite{hershey2021benefit, gemmeke2017audio}, a dataset of 79k audio files with strong labels available for download.

We used the Polyphonic Sound Detection Score (PSDS) Scenario 1 and 2~\cite{bilen2020framework}, event F1 (E-F1), and intersection F1 (I-F1)~\cite{mesaros2016metrics, giannoulis2013database} as evaluation metrics. Note that PSDS1 places more emphasis on the accuracy of event detection timing, whereas PSDS2 focuses more on that of class prediction. 

\begin{table}[t]
    \setlength\abovecaptionskip{0mm}
    \setlength\belowcaptionskip{0mm}
    \setlength\tabcolsep{1.6mm}
    \centering
    \caption{Quantitative results}
    \vspace{-0.4mm}
    \begin{tabular}{p{3.7cm}|cccc}
        \toprule
        Model & PSDS1 & PSDS2 & E-F1(\%)&I-F1(\%)\\
        \midrule
        CRNN baseline~\cite{DCASE2023Workshop} & 0.352 & 0.579 & 45.7 & 65.8 \\
        w/ Formula-SED (100k) & \textbf{0.405} & \textbf{0.641} & \textbf{49.6} & \textbf{72.3} \\
        w/ Audioset Strong (79k)~\cite{hershey2021benefit} & 0.387 & 0.618 & 47.7 & 70.7 \\
        \midrule
        Paderborn CRNN\cite{ebbers2022pre}& 0.262 & 0.506 & 34.9 & 57.3 \\
        w/ Formula-SED (100k) & 0.278 & 0.539 & 35.3 & 59.4\\
        w/ Audioset Strong (79k)~\cite{hershey2021benefit} & \textbf{0.355} & \textbf{0.622} & \textbf{43.9} & \textbf{67.8}  \\
        \bottomrule
    \end{tabular}
    \label{tab:quantitative}
    \vspace{-5mm}
\end{table}


\subsection{Quantitative Comparison}
\begin{figure}[t]
    \centering
    \vspace{-0.8mm}
    \includegraphics[width=0.9\hsize]{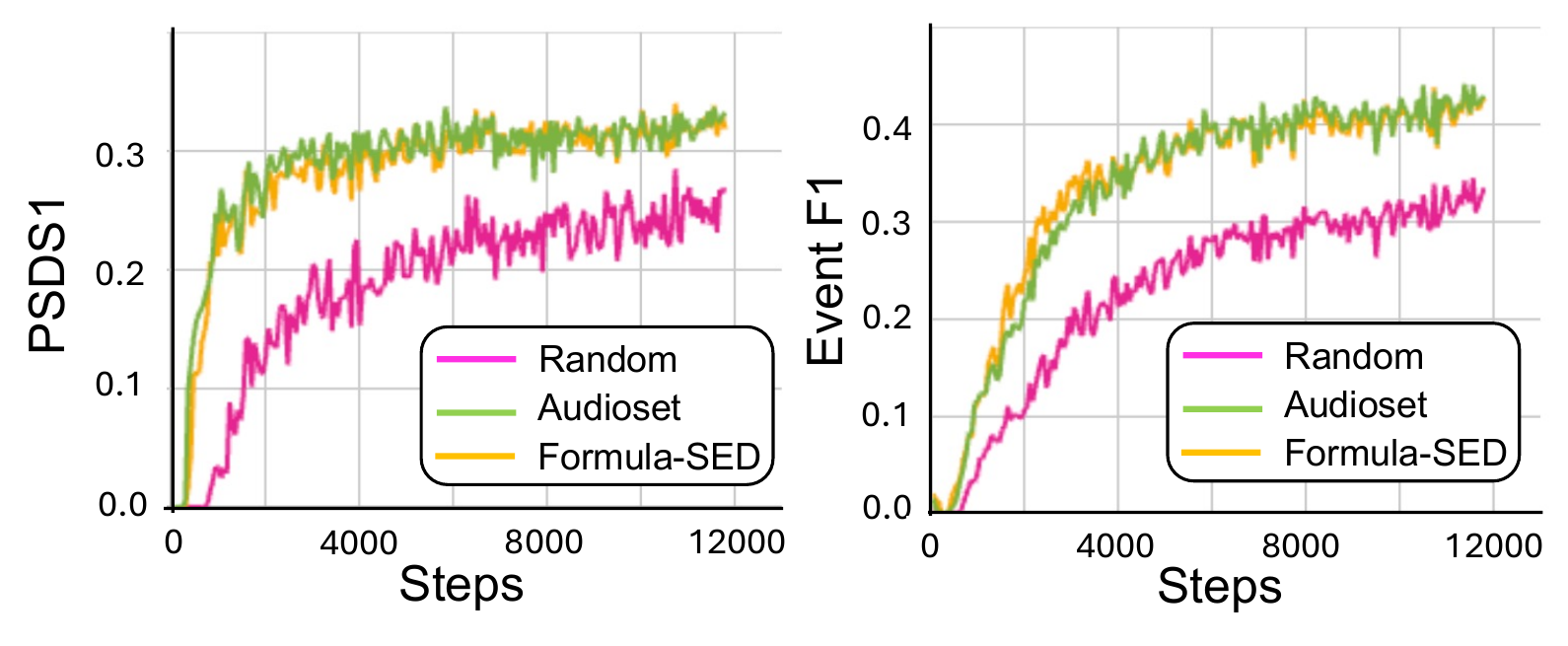}
    \vspace{-5mm}
    \caption{The training curve of the CRNN baseline~\cite{DCASE2023Workshop}.}
    \vspace{-3mm}
    \label{fig:training_curve}
\end{figure}

Table~\ref{tab:quantitative} shows the quantitative results of the three aforementioned pre-training methods. From these results, it can be observed that our proposed formula-driven method significantly improved the accuracy of both CRNN baseline and Paderborn CRNN on downstream tasks despite not using any real data. In the CRNN baseline, we can see that higher accuracy was recorded across all metrics compared to when a strongly labeled real dataset was used. Additionally, the learning curves in Fig.~\ref{fig:training_curve} demonstrate that our pre-training method effectively reduces the time required for convergence.

\begin{table}[t]
    \setlength\abovecaptionskip{0mm}
    \setlength\belowcaptionskip{0mm}
    \setlength\tabcolsep{1.6mm}
    \centering
    \caption{The impact of acoustic labels during pre-training}
    \begin{tabular}{p{3.2cm}|cccc}
        \toprule
        Model & PSDS1&PSDS2&E-F1(\%)&I-F1(\%)\\
        \midrule
        Baseline w/o pre-training & 0.352 & 0.579 & 45.7 & 65.8 \\
        Global F0 & 0.396 & 0.605 & 48.2 & 69.1 \\
        Local F0 & 0.384 & 0.612 & 48.9 & 70.3 \\
        Envelope sharpness & 0.361 & 0.601 & 48.9 & 70.9 \\
        Harmonic envelope & 0.391 & 0.628 & 49.3 & 71.1 \\
        Harnomic/Noise corr & 0.390 & 0.615 & 48.4 & 69.7 \\
        Noise distribution & 0.383 & 0.616 & 47.7 & 69.6 \\
        Reverb & 0.389 & 0.610 & 48.8 & 70.7 \\
        Ours & \textbf{0.405} & \textbf{0.641} & \textbf{49.6} & \textbf{72.3} \\
        \bottomrule
    \end{tabular}
    \label{tab:additional_experiments}
    \vspace{-4.5mm}
\end{table}

\subsection{Critical Components for Transferable Auditory Acquisition}
\begin{figure}[t]
    \centering
    \includegraphics[width=1.0\hsize]{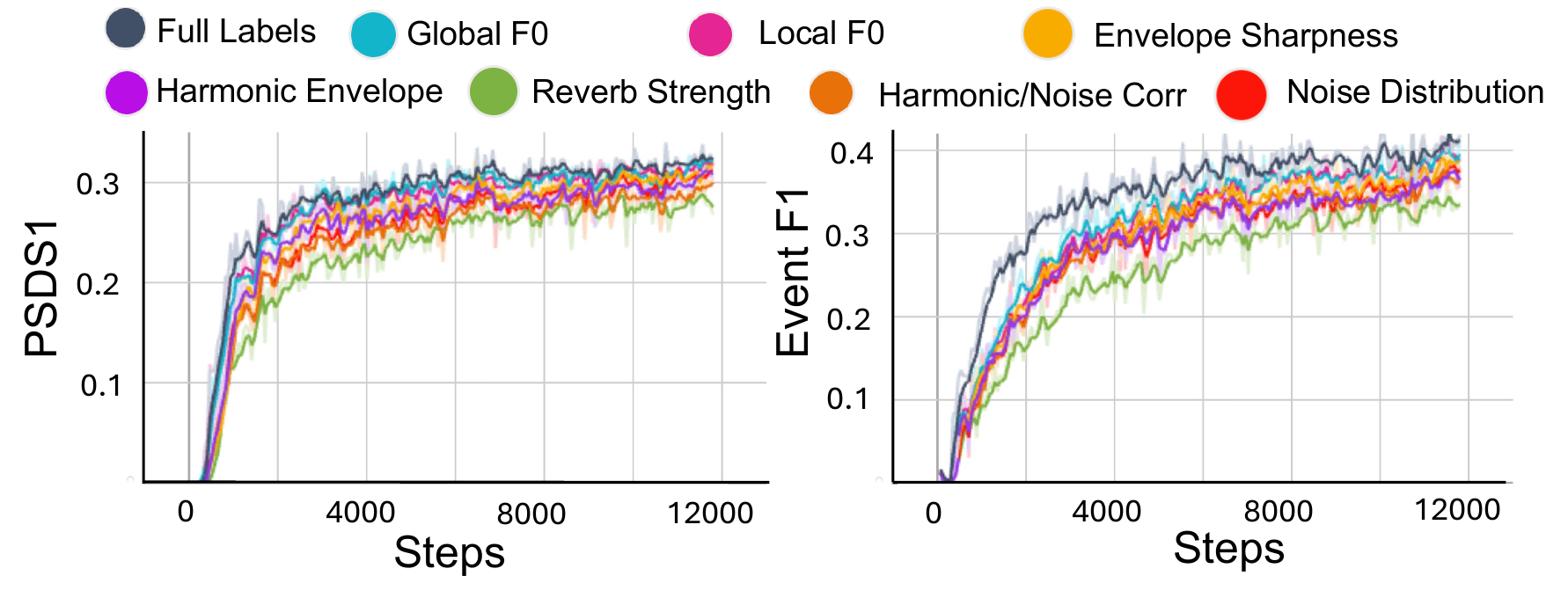}
    \vspace{-9mm}
    \caption{The impact of pre-training labels on fine-tuning performance.}
    \vspace{-2.5mm}
    \label{fig:training_curve_exp_psds1}
\end{figure}

Leveraging the advantage of constructing Formula-SED using a finite set of synthesis parameters, we conducted pre-training using a subset of these parameters as supervision to investigate which elements contribute to improving downstream task accuracy. The results are shown in Table~\ref{tab:additional_experiments}. Note that Global F0 includes the kernel, length scale, and variance hyperparameters related to the global fundamental frequency and that other rows also represent several hyperparameters related to one acoustic component. Additionally, Fig.~\ref{fig:training_curve_exp_psds1} shows a training curve under these settings. These results indicate that by predicting global/local frequency variations or harmonic envelope, the accuracy of downstream SED tasks is significantly improved. It can also be observed that labels related to noise and reverberation have a relatively weaker pre-training effect. 
Furthermore, from the two learning curves, it can be seen that the proposed method, which utilized all labels except reverberation, resulted in the fastest convergence during training. These results suggest the importance of considering various acoustic components during pre-training.

\subsection{The Impact of Pre-Training Dataset Size}
Since our proposed dataset is automatically generated based on mathematical formulas, it can easily scale in data quantity without incurring data collection costs or raising privacy and data ownership concerns. We compared the accuracy of the CRNN baseline by varying the dataset scale to 50k, 100k, and 1M. The results are shown in Table~\ref{tab:datasize}. These results confirm that the proposed dataset has a pre-training effect even at a smaller scale, such as 50k samples. Additionally, in the baseline CRNN, accuracy improved monotonically with the increase in the pre-training data scale. In the Paderborn CRNN, a correlation was observed between the pre-training data scale and downstream task accuracy for metrics that strictly evaluate detection timing, such as PSDS1 and Event F1. This suggests that the proposed pre-training method effectively utilizes the precise timestamped labels provided by our Formula-SED.

\begin{table}[t]
    \setlength\abovecaptionskip{0mm}
    \setlength\belowcaptionskip{0mm}
    \setlength\tabcolsep{1.4mm}
    \small
    \centering
    \caption{The impact of pre-training dataset size}
    \begin{tabular}{p{2.5cm}|l|cccc}
        \toprule
        Model & Size & PSDS1&PSDS2&E-F1(\%)&I-F1(\%)\\
        \midrule
        CRNN baseline & 50k & 0.380 & 0.620 & 49.4 & 70.8\\
        CRNN baseline & 100k & 0.405 & 0.641 & 49.6 & 72.3\\
        CRNN baseline & 1M & \textbf{0.420} & \textbf{0.653} & \textbf{51.4} & \textbf{72.8}\\
         \midrule
        Paderborn CRNN & 50k & 0.288 & \textbf{0.553} & 35.8 & \textbf{62.4}  \\
        Paderborn CRNN & 100k &  0.278 & 0.539 & 35.3 & 59.4\\
        Paderborn CRNN & 1M & \textbf{0.304} & 0.552 & \textbf{37.1} & 61.4\\
        \bottomrule
    \end{tabular}
    \label{tab:datasize}
    \vspace{-2mm}
\end{table}

\begin{table}[!t]
\setlength\abovecaptionskip{0mm}
\setlength\belowcaptionskip{0mm}
\setlength\tabcolsep{1.8mm}
\small
\centering
\caption{The Impact of Data Augmentation during pre-training}
\begin{tabular}{p{3cm}|cccc}
    \toprule
    Model & PSDS1&PSDS2&E-F1(\%)&I-F1(\%)\\
    \midrule
    Ours & 0.405 & 0.641 & 49.6 & 72.3 \\
    w/o Mixup~\cite{zhang2017mixup} &  0.376 & 0.625 & 49.1 & 72.5\\
    w/o time-shifting & 0.386 & 0.616 & 50.0 & 71.4 \\
    w/o time-warping~\cite{ebbers2022pre} & 0.384 & 0.618& 50.0 & 71.6\\
    w/o time-masking~\cite{park2019specaugment} & 0.397 & 0.618 & 49.2 & 70.9\\
    \bottomrule
\end{tabular}
\label{tab:data_augmentation}
\vspace{-4mm}
\end{table}

\subsection{The Impact of Data Augmentation}
As shown in Fig.~\ref{fig:data_samples}, our dataset differs from real data in terms of spectrogram appearance. 
To mitigate this domain gap, data augmentation has been reported as crucial for formula-driven supervised learning for computer vision tasks~\cite{nakamura2024scaling}.
To verify whether this trend holds, we present quantitative results when pre-training was conducted after individually removing each data augmentation in Table~\ref{tab:data_augmentation}. We can see that by utilizing the described data augmentation~\cite{park2019specaugment,ebbers2022pre,zhang2017mixup}, we consistently achieved high accuracy across many metrics.


\section{Conclusion}
In this paper, we proposed a supervised pre-training method utilizing our Formula-SED dataset, generated entirely without using real data.
The dataset is constructed through formula-driven acoustic signal synthesis, along with its corresponding synthesis parameters as labels.
Our fully synthesized dataset effectively addresses issues related to label noise, bias, and data ownership rights. In the SED task, the proposed pre-training method achieves both improved model accuracy and faster learning. These results demonstrate, for the first time, that auditory representations learned from mathematical formulas can be successfully transferred to real-world data.


\newpage

\bibliographystyle{IEEEtran}
\bibliography{IEEEfull}

\end{document}